\documentclass[
reprint,
amsmath,amssymb,
aps,
prl,
floatfix,
longbibliography,
]{revtex4-2}

\usepackage{comment}
\usepackage[normalem]{ulem}
\usepackage[T1]{fontenc}
\usepackage[utf8]{inputenc}
\usepackage[hidelinks]{hyperref}
\usepackage{physics}
\usepackage[dvipsnames]{xcolor}
\usepackage{graphicx}
\usepackage{dcolumn}
\usepackage{bm}

\usepackage{xcolor} 
\usepackage{soul} 
\usepackage{ulem}

\date{\today}

\begin{document}

\title{Speckle renewal, not bandwidth alone, mitigates kinetic Raman scattering at moderate bandwidth}
\author{P.-E. Masson-Laborde$^{1,2}$}
\email{paul-edouard.masson-laborde@cea.fr}
\author{G. Bouchard$^{1,2}$}
\author{R. Capdessus$^{1,2}$}
\author{C. Ruyer$^{1,2}$}
\author{A. Debayle$^{1,2}$}
\author{A. Fusaro$^{1,2}$}

\affiliation{$^1$CEA, DAM, DIF, F-91297 Arpajon, France}
\affiliation{$^2$Universit\'e Paris-Saclay, CEA, LMCE, 91680 Bruy\`eres-le-Ch\^atel, France}

\begin{abstract}
Broadband lasers are expected to mitigate laser--plasma instabilities by shortening the coherence time. Using two-dimensional particle-in-cell simulations of kinetic Raman scattering, we compare two broadband beams with identical power spectra but different speckle dynamics. At $2\%$ bandwidth, where Brillouin scattering is negligible, induced spatial incoherence (ISI) gives $34\%$ less Raman backscatter than a broadband beam focused through a random phase plate (RPP+BB), an ordering that persists with immobile ions. The two fields differ in their finite-time speckle contrast.
\end{abstract}

\maketitle

Burning-plasma experiments and target gains above unity at the National Ignition Facility have renewed interest in inertial confinement fusion (ICF) and inertial fusion energy (IFE) \cite{Lawson,Ignition,Kritcher2024}. Laser--plasma instabilities (LPI), however, still limit laser coupling through stimulated scattering and hot-electron production \cite{Craxton2015,Montgomery2016,Hurricane2023}. Stimulated Raman scattering (SRS) is a particular concern in direct-drive configurations \cite{Rosenberg2018,Solodov2022}. Random phase plates (RPP), smoothing by spectral dispersion, and induced spatial incoherence (ISI) were introduced to reduce such instabilities \cite{Kato1984,Skupsky1989,Lehmberg1983,Lehmberg1987}. A complementary route is to shorten the coherence time with incoherent broadband (BB) light, which finite-bandwidth theory has long shown to reduce parametric growth \cite{Thomson1974,ThomsonKruer1974,Thomson1975}. The idea is receiving renewed attention for IFE drivers, with recent simulations predicting reduced SRS, SBS, and two-plasmon decay \cite{Zhao2015,Zhao2017,Zhou2018,Bates2018,Follett2018,Ma2021,Liu2022,Luo2022,Zhao2023,Liu2024,Blackman2024,Zhao2025}.

High-energy broadband facilities such as Kunwu, Fourth-generation Laser for Ultrabroadband eXperiments (FLUX), and Petawatt High-Energy Laser for Heavy Ion Experiments (PHELIX) now allow this idea to be tested \cite{Eimerl2014,Dorrer2025,Gao2020a,Kanstein2025PHELIX}. With sub-percent bandwidth, experiments have already shown a strong reduction of backward stimulated Brillouin scattering (SBS) and improved coupling \cite{Wang2024,Lei2024}. The situation is less settled for SRS in the kinetic regime, where electron trapping, nonlinear frequency shifts, side loss, and inflationary amplification all modify the plasma response \cite{Vu2002, Yin2006, Masson2010, Yin2012, tacu2022, benisti2022}. Recent simulations indicate that moderate bandwidth need not suppress inflationary SRS \cite{Blackman2024}, and models with independent spatial and temporal randomness, as in RPP+BB below, find that a bandwidth below the growth rate can even enhance stimulated scattering~\cite{Tikhonchuk2026}.

The frequency spectrum, however, does not fully characterize a broadband laser field: how the spectral components are distributed across the focal spot also sets the spatiotemporal intensity fluctuations seen by the plasma. Spike Trains of Uneven Duration and Delay (STUD pulses) exploit this, combining temporal modulation with speckle scrambling to substantially reduce kinetic SRS~\cite{Afeyan2013STUD,Huller2013STUDSBS,Albright2014STUDSRS}. The same distinction separates two stochastic broadband architectures. When a broadband beam crosses a random phase plate (RPP+BB), all spectral components produce the same speckle pattern, whose brightness fluctuates in time. In ISI~\cite{Lehmberg1983,Lehmberg1987}, spatiotemporal decorrelation of the electric field is achieved through an echelon that introduces spatially varying time delays while also acting as a phase plate. This is equivalent to giving each spectral component its own phase plate, so that distinct speckle patterns evolve through temporal interference. Experiments have demonstrated Raman reduction with ISI~\cite{Obenschain1989}, and linear theory has identified an additional stabilizing effect of moving hot spots beyond bandwidth-induced detuning~\cite{Guzdar1991}. The consequences for nonlinear kinetic Raman scattering are less established. The two fields can share the same power spectrum and first-order coherence and still differ in their intensity statistics: RPP+BB has a heavier intensity tail than the nearly exponential ISI distribution~\cite{Odonnell1982}, but instantaneous statistics alone do not tell how long local intensity fluctuations persist.

Here, we show that identical bandwidths and coherence times do not imply identical kinetic Raman responses. Using two-dimensional particle-in-cell (PIC) simulations in a kinetic regime ($k_{\mathrm{EPW}}\lambda_{De}\simeq0.33$), we find that RPP+BB and ISI produce different Raman reflectivities at $2\%$ bandwidth, where SBS is negligible. The separation persists with immobile ions, so Brillouin-mediated pump depletion is not its necessary origin. We then show that the two fields, despite identical power spectra and first-order coherence, have different intensity covariances and finite-time speckle contrasts.


RPP+BB and ISI differ in how the phase-plate realization is assigned to the spectral components. For RPP+BB, all components share the same realization, so after factoring out the carrier $\omega_0$ the field separates into spatial and temporal factors,
\begin{equation}
\begin{aligned}
E_{\mathrm{RPP+BB}}(x,y,t)
&= E_s(x,y;\boldsymbol{\varphi})\,s(t),\\
s(t)
&=
\sum_{m=1}^{\mathcal{M}}
w_m\, e^{i(\phi_m-\Omega_m t)},
\end{aligned}
\label{eq:eq1}
\end{equation}

where $E_s(x,y;\boldsymbol{\varphi})$ is the complex speckle field produced by the phase plate, whose $N$ element phases, $[\boldsymbol{\varphi} \equiv {\varphi}_{n=1}^{N}]$, are drawn uniformly over $[0,2\pi]$~\cite{oudin2021,oudin2022}. $\Omega_m$ is the detuning of mode $m$ from the line center, $\phi_m$ its independent random spectral phase, and $w_m$ its amplitude, normalized through $\sum_m|w_m|^2=1$; $s(t)$ contains the broadband stochastic temporal modulation, leading to ``temporal speckles''. 
Here $\mathcal M$ is the number of modes, with $|w_m|^2=1/\mathcal M$ (flat power spectrum) \cite{Liu2022,Liu2024}. 
One mode is placed at a random position within each of the $\mathcal M$
equal intervals of width $\delta\Omega=\Delta\omega/\mathcal M$ that divide
the band.
Previous one-dimensional work suggests that, once the correlation time is matched, the detailed spectral shape has only a secondary effect \cite{Tikhonchuk2026}.
In RPP+BB all spectral components pass through the same phase plate. They
therefore produce the same speckle pattern, and the bandwidth only makes its
brightness fluctuate, as $|s(t)|^2$, on the coherence time $\tau_c$.

In the idealized ISI beam used here, each echelon has the same size as the phase plate element. A different random temporal signal is generated for each of these echelons following the previously described method. Mathematically, each spectral component produces its own speckle pattern: the phases
$\psi_{n,m}$ of the plate elements are drawn independently for every
mode~$m$,
\begin{equation}
E_{\rm ISI}(x,y,t)=\sum_{m=1}^{\mathcal M} w_m\,E_{s,m}(x,y;\psi)\,
e^{-i\Omega_m t},
\end{equation}
where $E_{s,m}$ is the speckle field obtained with the phases $\psi_{n,m}$,
$n=1,\dots,N$; the spectral phases $\phi_m$ are absorbed in $\psi_{n,m}$.
Because the patterns of different components are independent, the way they
interfere, and hence the position of the hot spots, changes after each
coherence time. The RPP+BB pattern breathes in place, whereas the ISI pattern
is renewed.

\begin{figure}[t]
    \centering
    \includegraphics[width=\columnwidth]{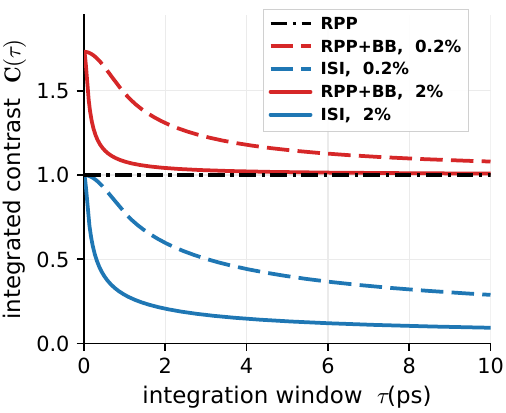}
\caption{Integrated contrast $C(\tau)$ against the averaging window $\tau$, at
$\Delta\omega/\omega_0=0.2\%$ and $2\%$. RPP+BB falls from $\sqrt3$ toward the RPP value $C=1$, ISI falls below it as $\tau^{-1/2}$. Curves from Eqs.~(\ref{eq:S}) and (\ref{eq:three}), flat
spectrum of width $\Delta\omega$.}
   \label{fig:fig1}
\end{figure}

The two broadband beams nevertheless have the same field coherence function
$g(\theta)=\langle E(t)E^*(t+\theta)\rangle/\langle|E(t)|^2\rangle$, where
$\langle\cdot\rangle$ denotes an average over the phase-plate and spectral
phases, but not the same intensity statistics over a finite time. Let
$u=I/\langle I\rangle$ be the normalized intensity,
$\kappa(\theta)=\langle u(t)u(t+\theta)\rangle-1$ its covariance, and
$u_\tau=\tau^{-1}\int_0^\tau\! u\,dt$ the intensity averaged over a window
$\tau$. The integrated contrast $C(\tau)=[\langle u_\tau^2\rangle/\langle
u_\tau\rangle^2-1]^{1/2}$ is then
\begin{equation}
C^2(\tau)=\frac{2}{\tau}\int_0^\tau\!\Big(1-\frac{\theta}{\tau}\Big)
\kappa(\theta)\,d\theta .
\label{eq:C2}
\end{equation}
as for any time-integrated speckle~\cite{goodman2007}. One has $\kappa=1$ for RPP, $\kappa=|g|^2$ for ISI and $\kappa=1+2|g|^2$ for
RPP+BB, so that the whole bandwidth dependence reduces to a single smoothing
factor
\begin{equation}
S(\tau)=\frac{2}{\tau}\int_0^\tau\!\Big(1-\frac{\theta}{\tau}\Big)
|g(\theta)|^2\,d\theta ,
\label{eq:S}
\end{equation}
and the three architectures follow in closed form,
\begin{equation}
C_{\mathrm{RPP}}=1,\quad
C_{\mathrm{RPP+BB}}=\sqrt{1+2S(\tau)},\quad
C_{\mathrm{ISI}}=\sqrt{S(\tau)} .
\label{eq:three}
\end{equation}
Since $S(0)=1$ and $S\simeq \tau_c/\tau$ for
$\tau\gg \tau_c$, with $\tau_c=2\int_0^\infty |g(\theta)|^2d\theta$ the integral coherence time,
RPP+BB decreases from $\sqrt3$ toward $1$ while ISI decreases from $1$ as
$\sqrt{\tau_c/\tau}$ [Fig.~\ref{fig:fig1}].
RPP+BB thus keeps a finite contrast for windows much longer than $\tau_c$, because its speckle pattern persists, whereas ISI keeps smoothing through speckle renewal. Identical first-order coherence therefore does not imply equivalent intensity smoothing, and $C(\tau)$ measures the difference without assuming any plasma response. The full derivation is given in the Supplemental Material~\cite{SM}.

The simulations are fully electromagnetic and two-dimensional, using \textsc{Smilei} \cite{Derouillat2018}. The target is a homogeneous CH plasma with $T_e=1\,\mathrm{keV}$, $T_i=0.3\,\mathrm{keV}$ and $n_e=0.05n_c$. The box size is $L_x=512\lambda_0$ along propagation and $L_y=256\lambda_0$ transversely, with $\lambda_0=527\,\mathrm{nm}$ and $n_c$ the critical density. The grid spacing is $\lambda_0/32$ with 128 particles per cell. The incident intensity spans $5\times10^{14}$--$3\times10^{15}\,\mathrm{W/cm^2}$.
The phase plate has either $N=16$ or $N=300$ elements and $f_\#=6$, giving a transverse speckle width $\simeq f_\#\lambda_0 = 6\lambda_0$, a longitudinal speckle length $10 f_\#^2\lambda_0 = 360\lambda_0$, and a focal spot spanning $\simeq 2Nf_\#\lambda_0 = 192\lambda_0$ for $N=16$. 

We consider $\Delta\omega/\omega_0=[0.2,0.8,2,10]\%$, i.e., $\tau_c=2\pi/\Delta\omega=[880,220,88,18]\,\mathrm{fs}$. The spectral mode density is the same in all cases, the number of modes increasing from $\mathcal{M}=20$ at $0.2\%$ to $\mathcal{M}=200$ at $2\%$.
We have $\delta\Omega\simeq10^{-4}\omega_0$, so that the recurrence time $2\pi/\delta\Omega\simeq17.6$~ps exceeds the simulated duration.

Simulations are performed with either mobile ions, allowing SRS and SBS to evolve self-consistently, or with immobile ions, which suppresses ion-acoustic dynamics. For each mobile-ion configuration, reflectivities are averaged over the $12\,\mathrm{ps}$ interaction and over two independent optical realizations with $N=16$ phase-plate elements. The reported spreads are their half-ranges, not statistical confidence intervals; the ordering of Raman reflectivities at $2\%$ bandwidth is preserved in both realizations. The immobile-ion calculations use $N=16$ and $N\simeq300$ phase-plate elements to test sensitivity to spatial speckle sampling. These configurations are distinct spatial tests, not independent realizations of the same optical ensemble.

\begin{figure*}[t]
    \centering
    \includegraphics[height=0.32\textwidth,trim=32 0 23 0,clip]{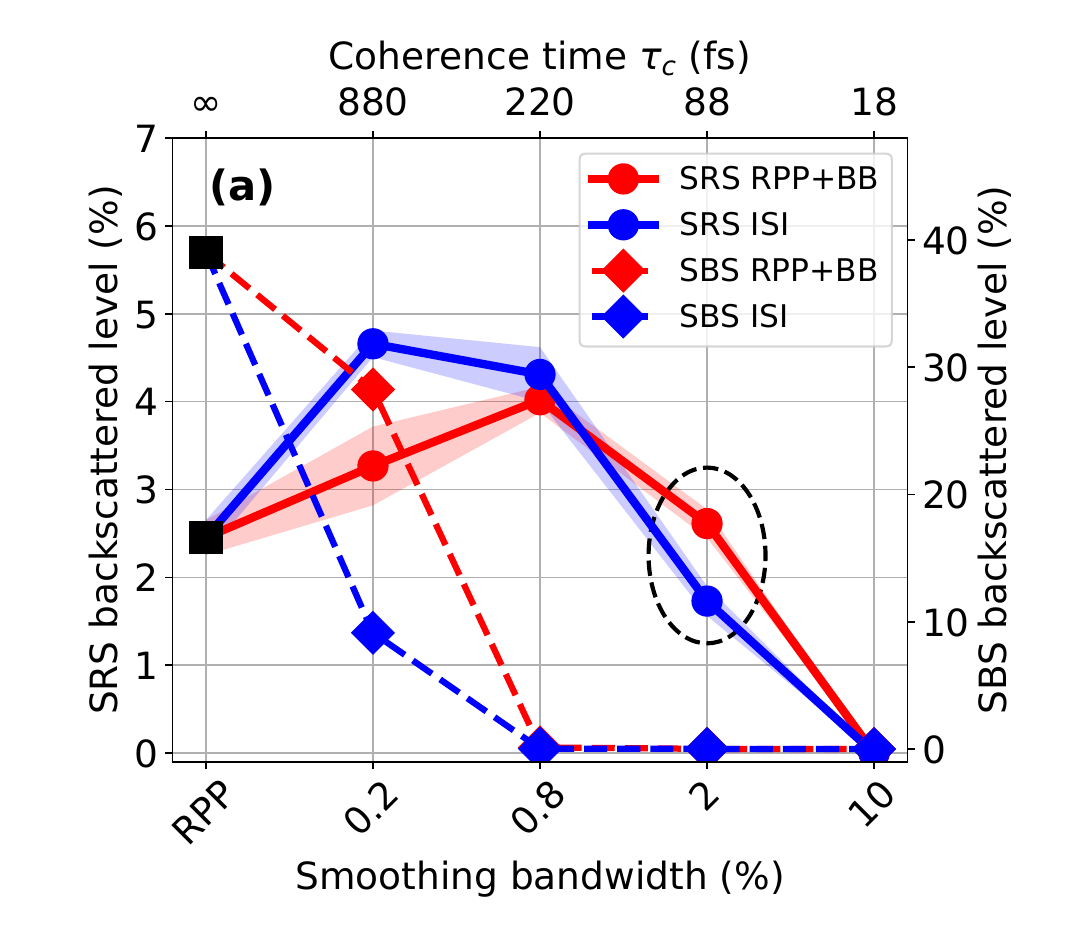}\hfill
    \includegraphics[height=0.32\textwidth,trim=21 0 73 0,clip]{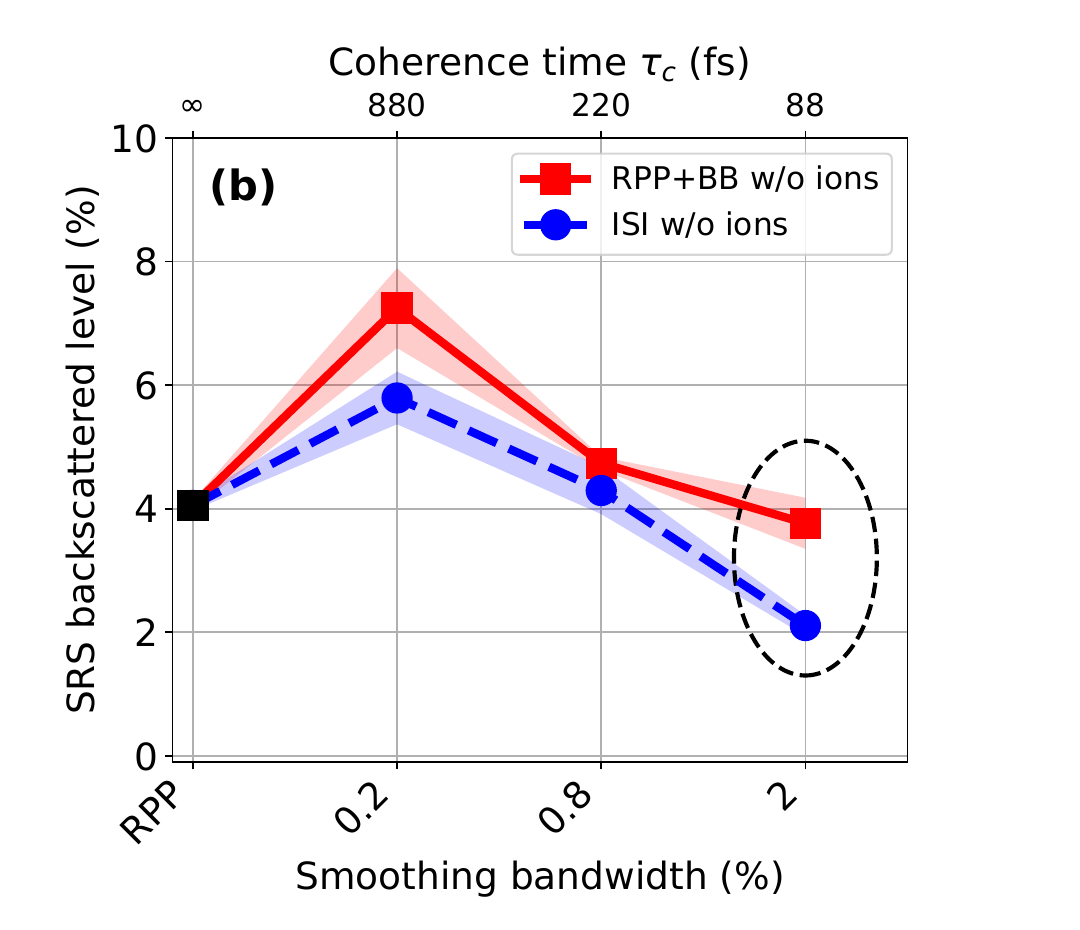}\hfill
    \includegraphics[height=0.32\textwidth,trim=32 0 23 0,clip]{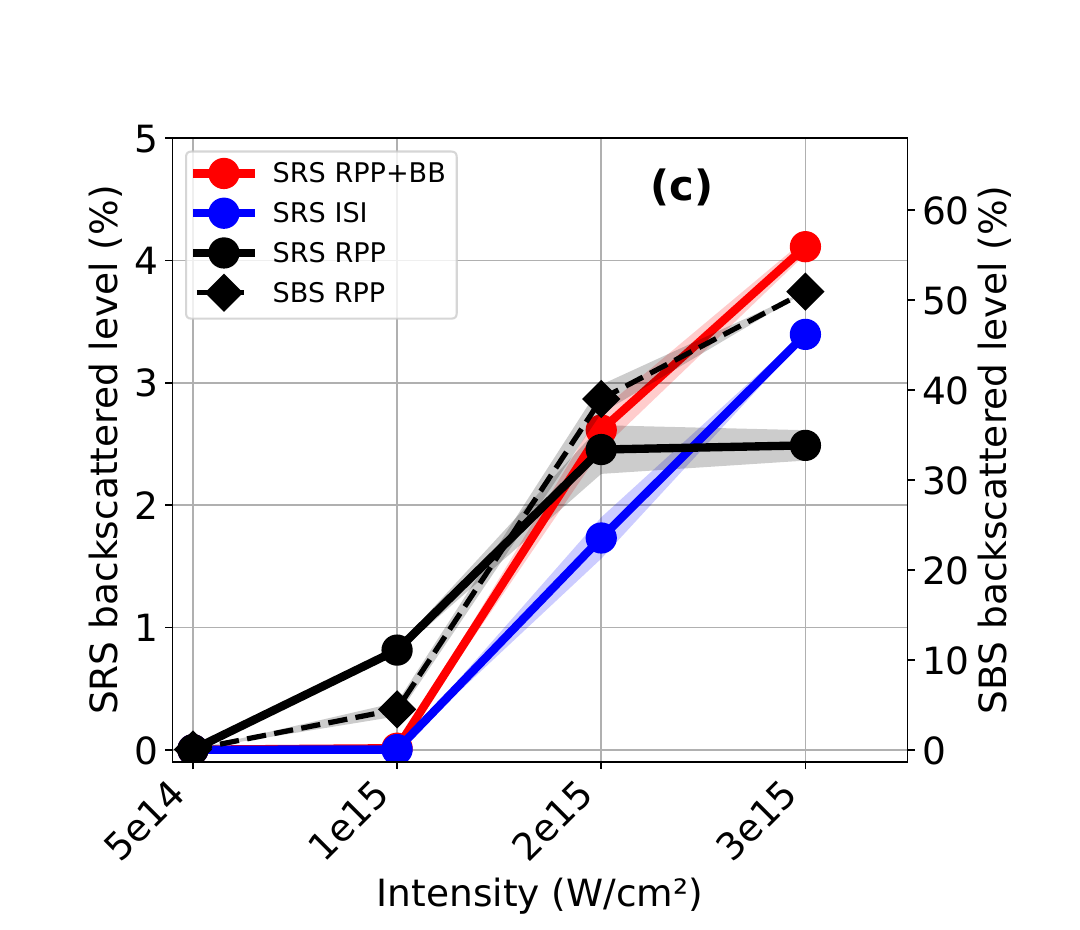}
\caption{SRS and SBS backscatter against bandwidth and laser intensity. (a) Time-averaged SRS (left axis) and SBS (right axis) with mobile ions, averaged over the $12\,\mathrm{ps}$ interaction and over two optical realizations; shaded bands give the half-range of the two. The dashed ellipse marks the $2\%$ point. (b) The same scan with immobile ions: average of the $N=16$ and $N\simeq300$ configurations; shaded bands show their half-range. (c) SRS (left axis) against incident intensity at $\Delta\omega/\omega_0=2\%$ with mobile ions, averaged as in (a), and SBS of bare RPP (dashed, right axis); SBS of the broadband beams is negligible and not shown.}
       \label{fig:fig2}
\end{figure*}
Figure~\ref{fig:fig2}(a) shows a clear difference between SBS and SRS as the bandwidth is increased for an intensity of  $2\times10^{15}$~W/cm$^2$. SBS drops rapidly, consistent with broadband detuning and with recent Kunwu and PHELIX measurements~\cite{Lei2024,Kanstein2025PHELIX}.

SRS is instead nonmonotonic. Sub-percent bandwidth increases the Raman reflectivity, and suppression is seen only at larger bandwidth. This trend has been observed in an experiment on PHELIX~\cite{Kanstein2025PHELIX}, and is expected when the bandwidth is below the Raman growth rate~\cite{Guzdar1991,Tikhonchuk2026}. At $10\%$ bandwidth, both beams suppress SRS through dephasing alone; this regime, beyond current facilities, is not considered further.

The two broadband beams also differ at a given bandwidth. At $0.2\%$, SBS reaches $28\%$ for RPP+BB against $9\%$ for ISI, while SRS is lower for RPP+BB ($3.27\pm0.45\%$ against $4.66\pm0.15\%$). SBS has disappeared at $0.8\%$, and at $2\%$ the SRS ordering reverses: ISI gives $1.73\pm0.17\%$ against $2.61\pm0.16\%$ for RPP+BB.
With mobile ions, SBS and SRS compete for the pump, so ISI, which suppresses SBS faster, leaves more energy for SRS at sub-percent bandwidth. The same competition limits the monochromatic reference. Bare RPP reflects $4.5\%$ of the light through SBS at $10^{15}$~W/cm$^2$, $39\%$ at $2\times10^{15}$ and $51\%$ at $3\times10^{15}$~W/cm$^2$ [Fig.~\ref{fig:fig2}(c)]; its SRS saturates, and both broadband beams exceed it at $3\times10^{15}$~W/cm$^2$. At $2\%$ bandwidth, SBS stays below $0.1\%$ for both broadband beams over the whole intensity scan.

Immobile ions [Fig.~\ref{fig:fig2}(b)] remove the Brillouin competition. At $2\%$ bandwidth, ISI still reflects less than RPP+BB ($2.11\pm0.17\%$ against $3.76\pm0.41\%$), so the lower ISI reflectivity does not require pump depletion by SBS. The sub-percent enhancement of RPP+BB also persists and is therefore not caused by SBS competition alone. ISI, too, stays at or above the monochromatic level at $0.2$ and $0.8\%$ ($5.8\%$ and $4.3\%$, against $4.05\%$) and falls below it only at $2\%$. There, the RPP+BB range overlaps the monochromatic reference ($2.61\pm0.16\%$ against $2.45\pm0.20\%$ with mobile ions, $3.76\pm0.41\%$ against $4.05\pm0.08\%$ with immobile ions), whereas ISI lies below both. In these conditions, broadband modulation of a fixed speckle pattern gives little net Raman reduction; decorrelating the pattern does. The intensity scan of Fig.~\ref{fig:fig2}(c) agrees: both broadband beams suppress SRS at $10^{15}\,\mathrm{W/cm^2}$, near the RPP threshold, where bandwidth alone dephases the growth, and ISI stays below RPP+BB at $2$ and $3\times10^{15}$~W/cm$^2$. We now relate this difference to the intensity statistics of the two fields.


We first consider the instantaneous statistics. For a fully developed RPP speckle field, $u$ is exponentially distributed \cite{goodman2007,Garnier1999,Rose1993}. In RPP+BB, the spatial and temporal complex amplitudes factorize. In the ideal fully developed-speckle limit, and for sufficiently many random spectral modes, the corresponding normalized spatial and temporal intensities are independent unit-mean exponential variables. Their product $u$ then follows the doubly scattered-speckle distribution previously derived for the product of two statistically independent speckle intensities~\cite{Odonnell1982},
\begin{equation}
P_{\mathrm{RPP+BB}}(u)
=
2K_0\!\left(2\sqrt{u}\right),
\label{eq:eq3}
\end{equation}
where $K_0$ is the modified Bessel function of the second kind. Equation~(\ref{eq:eq3}) therefore defines the ideal factorized-speckle limit, which has a substantially enhanced high-intensity tail. The ISI field is instead a sum of independent speckle fields, each circular complex Gaussian, and is therefore Gaussian itself~\cite{goodman2015}, so that $P_{\mathrm{ISI}}(u)=e^{-u}$ as for RPP.

\begin{figure*}[t]
    \centering
    \setlength{\tabcolsep}{4pt}
    \begin{tabular}{ccc}
        \includegraphics[width=0.32\textwidth]{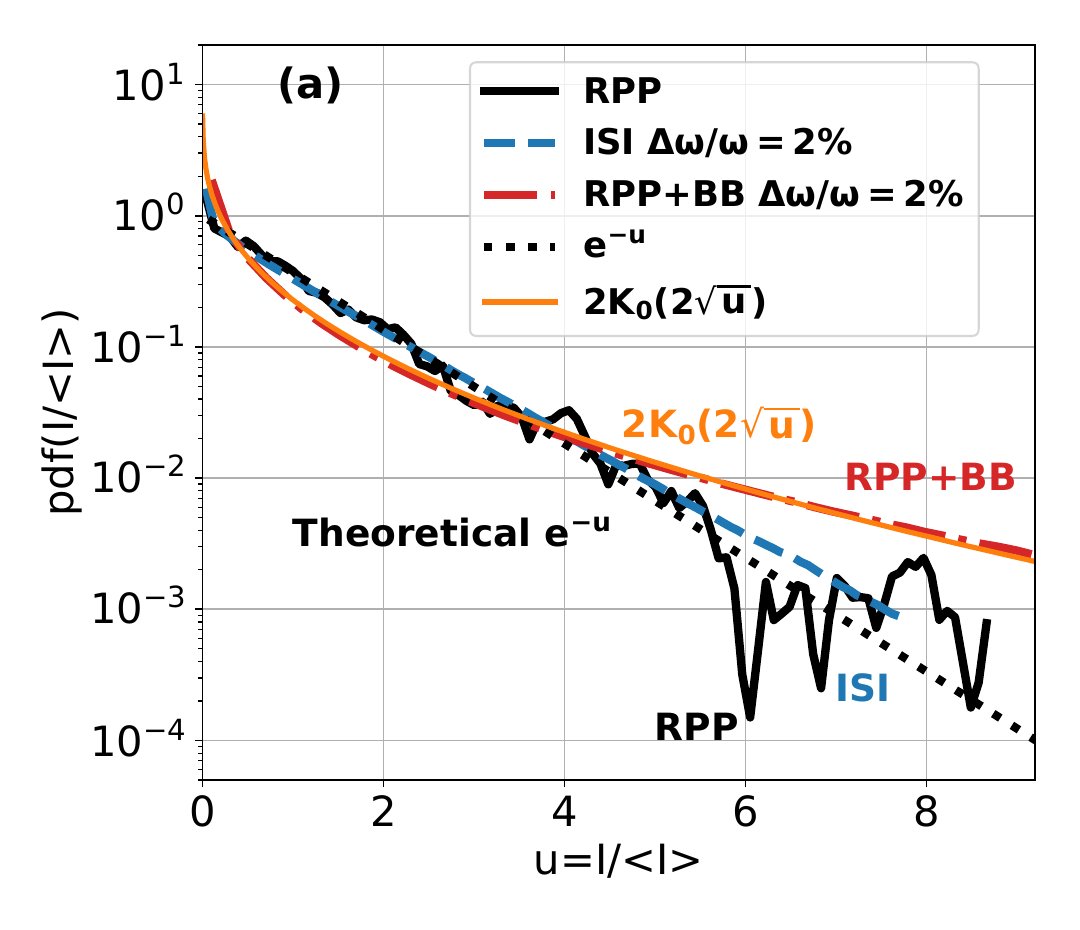}
        &
        \includegraphics[width=0.32\textwidth]{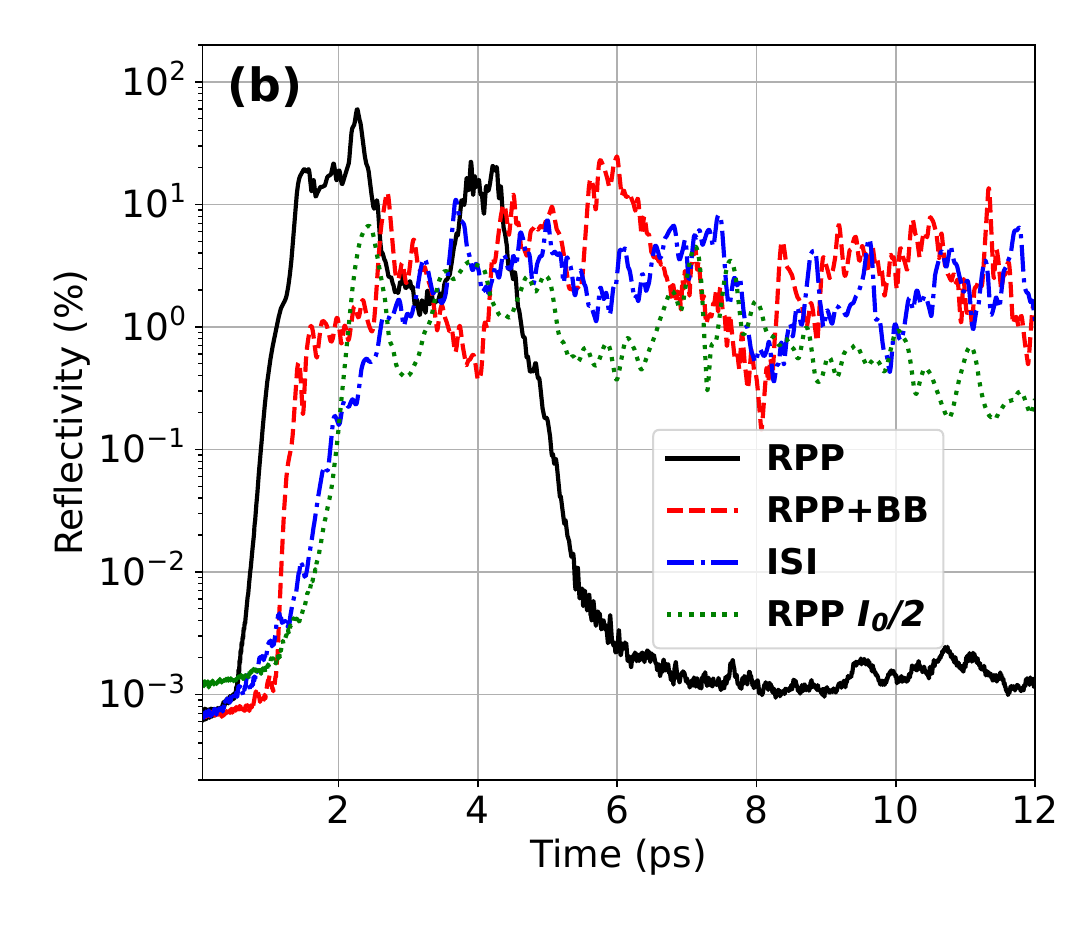}
        &
        \includegraphics[width=0.32\textwidth]{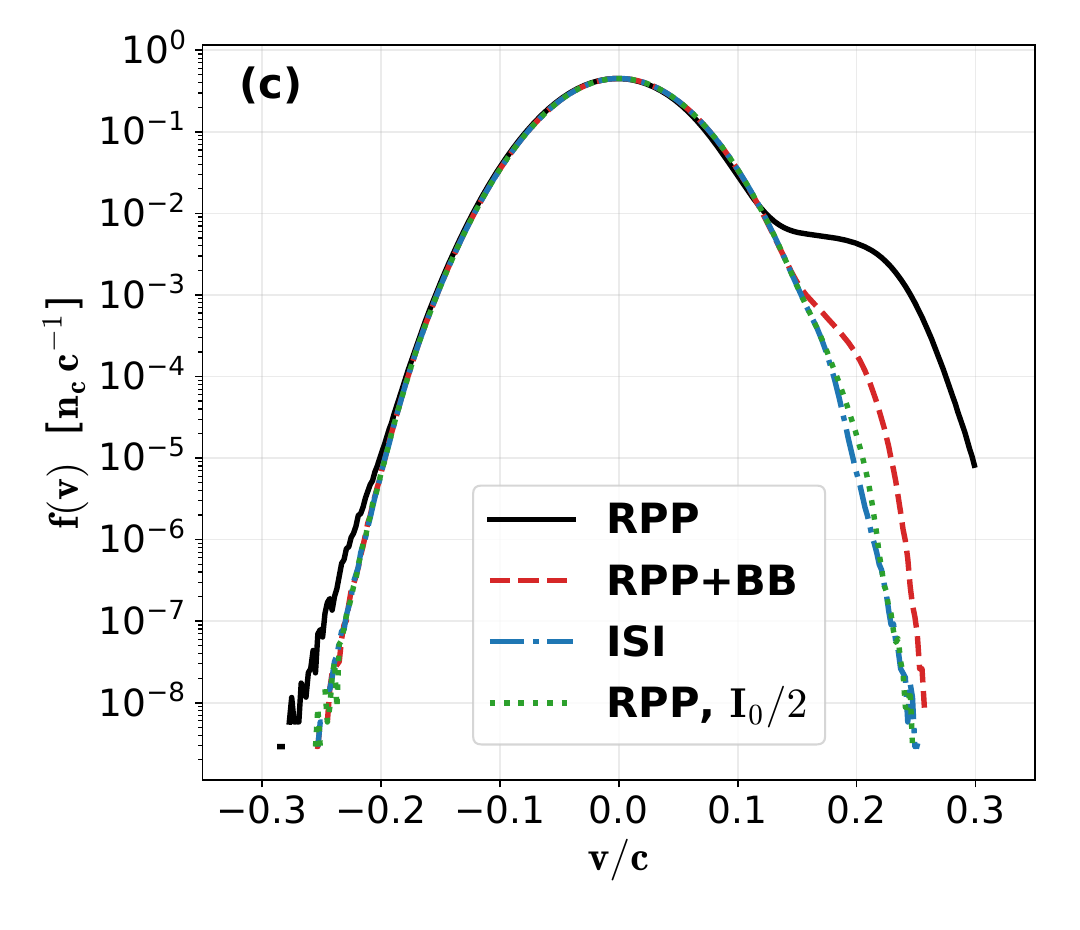}
    \end{tabular}
\caption{Laser-intensity statistics and nonlinear Raman dynamics. (a)
Probability density of the normalized intensity $u=I/\langle I\rangle$ at $\Delta\omega/\omega_0=2\%$, sampled at the plasma entrance over the full interaction; the exponential RPP law and the factorized limit $2K_0(2\sqrt u)$ are shown for reference. (b) SRS reflectivity over the $12\,\mathrm{ps}$ interaction at $I=2\times10^{15}\,\mathrm{W/cm^2}$ and $\Delta\omega/\omega_0=2\%$; the dotted curve is bare RPP at half that intensity. (c) Electron velocity distribution $f(v_x)$ along the propagation direction at $t=2\,\mathrm{ps}$, same cases as (b).}
     \label{fig:fig3}
\end{figure*}

Figure~\ref{fig:fig3}(a) shows the incident intensity PDFs sampled at the plasma entrance over the full interaction. ISI closely follows the exponential law, whereas RPP+BB at $2\%$ develops a heavy tail, approaching but not reaching the $2K_0(2\sqrt u)$
limit. The two distributions cross near $u\simeq3.2$; above it, RPP+BB samples bright fluctuations more often, with $\Pr(u>5)$ about three
times larger than for ISI. Such fluctuations may favor nonlinear amplification in the kinetic regime, where electron trapping makes the response sensitive to intense local events. However, an instantaneous PDF does not specify how long such fluctuations persist. ISI and monochromatic RPP, for example, have nearly identical exponential intensity distributions but different Raman responses; bandwidth-induced dephasing also contributes to this comparison. The finite-time statistics introduced above therefore complement the instantaneous PDF by characterizing the temporal persistence of intensity fluctuations and the renewal of speckles.

The second moments of the PDFs, Eq.~(\ref{eq:eq3}) and $P_{\mathrm{ISI}}=e^{-u}$, give the $(\tau\to0)$ limits of Eq.~(\ref{eq:three}), $C_{\rm RPP+BB}=\sqrt3$ and $C_{\rm ISI}=1$. This explains the larger instantaneous contrast of RPP+BB, but an instantaneous intensity PDF by itself cannot distinguish ISI from RPP. Their temporal intensity statistics differ because the ISI pattern decorrelates; their different contrasts must also be taken into account when interpreting the Raman comparison. For any finite window, Eq.~(\ref{eq:three}) gives $C_{\mathrm{ISI}}<C_{\mathrm{RPP}}<C_{\mathrm{RPP+BB}}$ since $0<S<1$.
Direct evaluation of the ISI contrast from the incident fields over the full $12\,\mathrm{ps}$ interaction gives $C_{\rm ISI}\simeq0.288$ at $0.2\%$ and $0.09$ at $2\%$ bandwidth, against $0.283$ and $0.091$ from Eqs.~(\ref{eq:S}) and (\ref{eq:three}) evaluated with the discrete spectrum used here.

The finite-time contrast measures the pump fluctuations that remain over the Raman amplification time. In the initial exponential phase of Fig.~\ref{fig:fig3}(b), where the reflectivity rises from $10^{-3}\%$ to $1\%$, RPP and RPP+BB grow approximately as $R\propto\exp(2t/\tau_g)$ with $\tau_g\simeq0.16$~ps, whereas ISI grows more slowly, $\tau_g\simeq0.5$~ps, in both optical realizations. At $2\%$ bandwidth, Eq.~(\ref{eq:three}) gives $C_{\rm ISI}\simeq0.66$ when evaluated over the RPP growth time, compared with unity for RPP and a larger value for RPP+BB. Thus, at this intermediate bandwidth, ISI already exhibits appreciable temporal smoothing over a timescale comparable to the observed Raman amplification, while RPP+BB retains a persistent spatial contribution.

The integrated contrast is not a predictive model for nonlinear Raman gain, which also depends on the instantaneous intensity distribution, plasma-wave dynamics, and electron trapping, and the present comparison does not separate the roles of the PDF and of temporal persistence. The reduced finite-time fluctuations of ISI are nevertheless consistent with its slower initial growth and lower reflectivity. Moving hot spots were known to add stabilization in linear theory~\cite{Guzdar1991}; the present simulations show that the difference between the two architectures persists in the nonlinear kinetic regime. For SBS, whose measured growth time is longer ($\tau_g\simeq0.57$~ps for bare RPP), temporal smoothing is effective at smaller bandwidths, consistent with Fig.~\ref{fig:fig2}(a).

The time traces in Fig.~\ref{fig:fig3}(b) show the same trend. Bare RPP shows a large Raman burst, quenched after $\simeq4$~ps, consistent with the growth of SBS, which reflects $39\%$ of the light at this intensity [Fig.~\ref{fig:fig2}(c)]. The stochastic temporal power modulations of the broadband beam drive strong bursts of SRS, which are substantially reduced with ISI owing to its spatiotemporal smoothing that reduced the power fluctuations. The stronger and more persistent RPP+BB response is consistent with the frozen spatial component of its intensity fluctuations.

The electron distributions in Fig.~\ref{fig:fig3}(c) show the kinetic signature of the Raman-driven plasma waves. All three configurations flatten
the distribution above the plasma-wave phase velocity
$v_\varphi\simeq0.153\,c$, and the extent of the suprathermal shoulder
follows the ordering of the reflectivities in Fig.~\ref{fig:fig3}(b): largest
for bare RPP, then RPP+BB, then ISI. The ISI distribution at
$2\times10^{15}\,\mathrm{W/cm^2}$ lies close to that of bare RPP at
$10^{15}\,\mathrm{W/cm^2}$, and the two start together and grow at similar rates [Fig.~\ref{fig:fig3}(b)]. During this first Raman burst, the ISI response is therefore close to that of a
monochromatic beam driven at half the intensity. Later, the ISI reflectivity grows above it, and its $12$-ps average is twice as large ($1.73\%$ against $0.81\%$) [Fig.~\ref{fig:fig2}(c)]. At $10^{15}$~W/cm$^2$, SBS takes only $4.5\%$ of the RPP light, so this comparison is little affected by Brillouin competition. 

Identical power spectra and coherence times therefore do not guarantee equivalent mitigation of kinetic Raman scattering. At moderate bandwidth, where Brillouin scattering is negligible, ISI produces substantially less Raman backscatter than broadband modulation of a fixed speckle pattern, an ordering that persists with immobile ions. The two fields differ in their finite-time intensity statistics: RPP+BB keeps a persistent spatial contrast, whereas ISI smooths local fluctuations through speckle renewal. The respective roles of the intensity distribution and of temporal persistence are not isolated here, but the measured growth histories are consistent with an additional role for speckle renewal. Broadband laser performance cannot be characterized by bandwidth alone: the spatiotemporal organization of the laser fluctuations is an additional degree of freedom for controlling kinetic laser--plasma instabilities.

This work was performed in the framework of the TARANIS project supported  by the French government as part of France 2030 (AAP Réacteurs Nucléaires Innovants - \#DOS0237678/00).

This project was provided with computer and storage resources by GENCI at TGCC, thanks to the allocations A0190515693 and A0210515693 on the supercomputer Joliot Curie's ROME partition.
This work has been done under the auspices of Commissariat à l’Energie Atomique (CEA), and some of the simulations were performed using high performance computing resources at Centre de Calcul pour la Recherche et la Technologie and CEA/Tera.

\bibliographystyle{apsrev4-2}
\bibliography{biblio}

\end{document}